\documentstyle[pre, floats, psfig, aps]{revtex}
\begin{document}
    \draft
\title{A self-regulated model of galactic spiral structure formation}
\author{
Daniel Cartin$^{1}$,
Gaurav Khanna$^{2}$}
\address{1. 89 Christopher Street, Apt. D,\\
Lodi NJ 07644}
\address{2. Natural Science Division,\\
Southampton College of Long Island University,\\
Southampton NY 11968.}

\date{\today}

\maketitle
\begin{abstract}
    
The presence of spiral structure in isolated galaxies is a problem that has
only been partially explained by theoretical models. Because the rate and
pattern of star formation in the disk must depend only on mechanisms internal
to the disk, we may think of the spiral galaxy as a self-regulated system
far from equilibrium. This paper uses this idea to look at a
reaction-diffusion model for the formation of spiral structures in certain
types of galaxies. In numerical runs of the model, spiral structure forms and
persists over several revolutions of the disk, but eventually dies out.

\end{abstract}


\section{Introduction}
\label{intro}

The problem of how spiral structures form in galaxies is one 
that has often been studied (for a general overview of the subject,
see \cite{bin87}), and can be divided into two aspects.
The first is that of temporary structures most likely caused by gravitational perturbations 
from passing galaxies or an asymmetric halo, or disk material having an initial velocity 
relative to the local standard of rest from formation processes. Here the 
phenomenon seems to be related to density waves \cite{lin64}, 
quasi-stable modes in the gravitational potential of the disk
which offer a good description of grand design spirals, where the 
arms are well-defined and has a high degree of symmetry.
However, this paper is devoted to the second part of the problem, namely,
that of spiral structures in ``isolated'' galaxies, i.e. where we 
ignore influences from outside the disk and consider only internal 
(and recurring) mechanisms. These galaxies can be represented by 
flocculent\footnote{The division of spiral galaxies into grand 
design and flocculent is based on the arm classification scheme of 
Elmegreen and Elmegreen \cite{elm-elm82}.} spirals,
so called because of the fleecy appearance of 
their many short and asymmetric spiral arms. Because these spirals
are seen in blue light, but not red \cite{elm-elm84} -- suggesting the 
arms are composed of younger, bluer stars, while the older, redder stars are more 
evenly distributed across the disk --
they must be an artifact of star formation itself, and are not primarily
density waves. Our attitude here is to consider structure formed by 
the processes of star formation and neglect gravitational influences.
This should apply equally to all 
spirals, although for grand design spirals, we would expect it to 
act with density waves arising from other sources.

Most of the theoretical work done in star formation 
processes deals with the solar neighborhood (those 
stars in the vicinity of our Sun) as opposed to structures on a 
galactic scale \cite{sho-fer95}. Other models used to study these aspects of star 
formation and spiral structure, such as those built on propagating star
formation \cite{sei-ger82}, either greatly simplify the physics involved or
need finely tuned parameters to match observations. In this paper,
we give a model of spiral structure in isolated 
galaxies based on the idea of star formation as part of a network of 
self-regulated and autocatalyzed reactions \cite{car00,CFKS}\footnote{This work, is
different from the work \cite{CFKS} done in the past, because it uses a modified model. 
In particular, the models in \cite{CFKS} had a subtle flaw in normalizations and were also
missing some physically important reaction terms. See \cite{car00} for a full discussion of 
this.}.  Recently, there has been a great
deal of both theoretical and experimental work studying how non-equilibrium
systems in chemistry and biology produce patterns in both time and
space. This has included looking at both the organic -- such as bacterial
colonies \cite{ben93}, the differentiation of cell
types \cite{kau93}, and the formation of
embryonic structure in multicellular organisms \cite{kau93} 
-- as well as the inorganic -- the BZ chemical reaction \cite{zai70}, diffusion
limited aggregation \cite{ben90}, and self-organized
critical systems \cite{bak87}. There have been many
successes in reproducing patterns in the
laboratory, and these models typically share both partial differential
equations and discrete elements such as cellular automata. However, most of this
work in non-equilibrium systems has been after the main work on the
formation of structures in astronomy.

Thus one might seek to apply this
work to patterns seen in spiral structures. Because we are studying the process in isolated
galaxies, we know that the formation must be caused by events within the disk,
rather than by the actions of outside players. The isolated galactic spiral
is far from equilibrium -- there is differentiation of material
into stars and clouds of gases whose distribution varies over space and time. 
In addition, star formation happens at a constant rate\footnote{This 
is true in observed galaxies up to a factor of two; see Sandage 
\cite{san86} for how the star formation rate in different types of 
spirals changes with time.}, as averaged across the 
disk. This is a clue that the process is regulated by a feedback
loop to maintain this constancy (for evidence of this mechanism, see 
\cite{bur90}).
These characteristics are
shared by other types of non-equilibrium systems. Below, we list the
predominant features that these kinds of networks of reactions have in
common, along with examples of the same behavior in galactic disks.

\begin{itemize}

\item {\bf Steady state system} \quad There is a slow (relative to the
dynamical time scales) and steady flow of energy, and perhaps matter running
through the disk. In spiral galaxies, star formation proceeds at a constant rate,
averaged over the disk, for time scales on the order of 10$^{10}$ years. The
fact that this is greater than the time scales of the actual star formation
process (10$^7$ years) implies that the slow and steady rate is
regulated by feedback mechanisms.

\item {\bf Non-equilibrium system} \quad The steady state is far from
thermodynamic equilibrium, and there is a coexistence of several species or
phases of matter, which exchange matter and energy among themselves through
closed cycles. The galactic disk is not a uniformly dense clump of material
at thermal equilibrium, but instead is divided into gases at different
temperatures and stars of various mass. These species exchange matter:
for example, massive stars supernova to form warm gas, which can cool and then condense
into new stars.

\item {\bf Feedback mechanisms} \quad The rates at which material flows
around these cycles are governed by feedback loops that have arisen during the
organization of the system into the steady state. An example of this is
suggested by Parravano and collaborators \cite{man-par91} which explains 
how the average pressure in the interstellar medium (ISM) is 
maintained. They argue that there are two phases, the warm gas of
the ambient phase, and the cooler gas of the condensed phase, with a
phase boundary in the pressure-temperature plane. Ultraviolet radiation from
the supernovae of massive stars heats the gas, which prevents the condensation
of newer stars, so the supernova rate goes down, allowing new massive stars
to form (and so the supernova rate will increase again). This feedback
mechanism keeps the gas on the phase boundary.

\item {\bf Autocatalytic reaction networks} \quad Any
substances that serve as catalysts or repressors of reactions
in the network are themselves produced by reactions inside the network.
Suppose we look at the condensation of giant molecular clouds (GMCs). This
is catalyzed by dust grains produced by cool giant stars, shielding the clouds
and providing sites for molecular binding, and carbon and oxygen, which may cool
the clouds by radiation from the rotational modes of CO molecules. The
condensation is inhibited by ultraviolet radiation from massive stars, as
described by the Parravano process mentioned above.

\item {\bf Separation in space} \quad There may be spatial segregation of the 
different phases or materials in the cycles. This occurs when the inhibitory
and catalytic influences propagate over different distance scales. At the
smallest scale, this means the production of certain substances may be subject
to refractory periods -- once production has occurred in a local region, it
will not be repeated there for some period of time. For the influences in
the process of GMC condensation, dust grains, carbon and oxygen propagate
only over distances of about 100 parsecs (how far supernovae and massive stars
can spread their products) while UV radiation can travel over much of the
galactic disk.

\end{itemize}

As can be seen, there is evidence that we can think of spiral structure
in isolated galaxies to be a product of a self-organized, autocatalyzed
network of reactions in the star formation process. Given a system with the
characteristics listed above, there are models which can describe the spatial
structure, the most typical of which is the reaction-diffusion model
\cite{kau93,tur52}, and we develop a model along these lines for spiral 
structure in isolated galaxies.

\section{Theoretical aspects}

\subsection{One-zone model}
\label{one-zone}

We briefly outline the processes necessary to start star formation.
The cold clouds of the GMCs condense out of the interstellar medium (ISM), forming
distributions of gas and dust that are apparently scale-invariant. As
mentioned above, the condensation is helped along by the actions of dust,
carbon and oxygen, while it can be impeded by UV radiation from massive stars.
The Parravano process may place some limitations on the amount of
condensation. Typical time scales for the inhibition are about $10^7$ years,
the average life span of massive stars, after which the supernova (SN) rate and
the UV radiation flux will die off. Once the GMCs start to condense, then
their cores may collapse to form stars. This collapsing is brought on by
shockwaves from supernovae or HII regions (we neglect any impact that 
density waves may have in these collapses), and so have the same length scale
as the propagation of dust by supernovae, or about 100 parsecs. Once the
stars are formed, they can inhibit the infall of gas by the stellar wind or
UV radiation produced by the star \cite{ada87}. These effects
occur on short length scales, about the size of one cloud complex, and
reduce the star forming efficiency of the clouds down to a few percent.

We must take the processes occurring in the galactic disk and abstract
them to produce a viable mathematical model. One way we can simplify the
system is to take the continuous spectrum of star types and break it up to
those massive stars that can supernova (and thus provide matter and radiation
back into the system) and those lighter stars that cannot. We will be
neglecting the fact that these lighter stars can return matter to the ISM, so
in our model, they will simply act as a matter sink. There is also
matter exchange between the cold gas of the
GMCs and the warmer ambient gas, due to heating and cooling. We can summarize
the flows of material in Fig. 1, and the various material and 
energy components are given in Table \ref{table1}. Now we give a brief 
description of the physical processes we include in the model.

\begin{itemize}
    
\item Cloud destruction $cs \to g, cs \to s$: Because of the presence of
massive stars and their stellar winds and SNe produced shocks, there will be
a mechanism of cloud distruption. Some of
this pressure will trigger star formation $(cs \to s)$, but the eventual
result will be the disruption of the cloud into warm gas $(cs \to g)$.
Typically, the efficiency of star formation is around a few percent.

\item Cloud-cloud collisions $c^2 \to g, c^2 \to s$: Another source of
pressure is the collision of clouds, which will have the same types of effects
as the star-induced cloud destruction mentioned above. Again, the 
rate of star formation is a few percent.

\item Mass infall to the disk $\delta$: Galaxies are believed to be formed
from the condensation of matter from a spherical halo into a disk, and so
there is certainly the possibility that there is a continuing flow of matter.
It is believed that the rate is enough to replenish the material in the disk
in a time span of billions of years (see Section 4.3 of 
Larson~\cite{lar92}).

\item Direct cloud destruction by stars $cs \to g$: This represents 
the effects of stars, such as stellar winds, which come directly from 
the massive stars, as opposed to radiation and shockwaves, which 
might travel some distance. The main physical action behind this term is the
ionization and champagne flows of stars formed inside the cloud~\cite{ten82}.

\item UV radiation, shockwave production $s \to r, s \to h$: The 
sources of UV radiation and shockwaves are from SN events from massive stars. These
effects are more long range, although shocks will travel only about 100 pc, while
radiation can traverse the entire galaxy.

\item Damping terms $(c + g) r \to r, (c + g) h \to h$: Because the
energy carried by UV radiation and shockwaves will be dissipated by the interaction 
with matter -- both warm and cold -- we include a damping term.

\item Cooling term $g^{2} / r \to c$: UV radiation will act as a thermostat, since
warm gas is less likely to cool in an environment with a high radiation 
density. In the results presented here, we consider cooling which is 
inversely proportional to the radiation density.
   
\item Cloud destruction by UV radiation and shockwaves $cr \to g, ch \to
s$:  These effects are more long-range than the direct cloud destruction used
previously. Note that ultraviolet radiation will ionize the clouds into warm
gas, as does the destruction induced by massive stars, but the pressure due to
shockwaves will initiate new star formation.

\end{itemize}

Now, we take this model of the material flows in the galactic disk,
and write them in a system of equations using the law of mass 
action -- that is, the change in time of the output quantity is given 
by the product of the densities of the inputs, with a constant 
parameter giving the rate of the reaction. Thus, given the processes 
outlined above, we have
\begin{mathletters}
\label{eqns-one-zone}
\begin{eqnarray}
\frac{d c}{d t} &=& \frac{\alpha_1 g^2}{r} - \alpha_2 cr - \beta_1 ch - \gamma cs -
	\epsilon_{1} c^2\\
\frac{d g}{d t} &=& - \frac{\alpha_1 g^2}{r} + \alpha_2 cr + s + \gamma cs + \epsilon_{2} c^2
	+ \delta \\
\frac{d s}{d t} &=& \beta_{2} ch + \epsilon_{3} c^2 - s\\
\frac{d r}{d t} &=& \eta_1 s - \phi_{1} (c + g) r \\
\frac{d h}{d t} &=& \eta_2 s - \phi_{2} (c + g) h \\
\frac{d d}{d t} &=& \beta_3 ch + \epsilon_4 c^2
\end{eqnarray}
\end{mathletters}
where the parameters of the equations $\alpha_1, \alpha_2, \cdots$
and their ranges are given in Table \ref{table2}. To choose the parameters, we 
use the lifetime of a typical massive star, $10^7$ years, as the 
dimension of time, and the units of mass and energy to be those 
appropriate for each components -- for example, we choose the mass 
unit for warm gas $g$ to be one hydrogen atom per cubic centimeter. 
Then each parameter is the rate of the reaction at the mass and energy 
densities we select, e.g. since the mass flow into the galaxy is 
estimated to completely replace the current material in $10^{10}$ 
years, we choose the mass inflow constant $\delta \sim 0.001$. For a 
fuller discussion of the choices made, see \cite{car00}. Finally, note that conversation of 
matter implies $\beta_1 = \beta_2 + \beta_3$ and $\epsilon_1 = 
\epsilon_2 + \epsilon_3 + \epsilon_4$.

As mentioned above, only massive stars are adding material to the ISM,
through the mechanism of supernovae -- we are ignoring the fact that
light stars add material via evaporation.  There are a few other
simplifications which have been used to arrive at these equations,
such as neglecting the impact of catalyzers to GMC condensation such
as dust and carbon to avoid parameters that depend on metal
concentrations.  
Also, we noted previously that our cooling term is
inversely proportional to the radiation density. This gives the simple
result that cooling is faster when there is less radiation, but this is
certainly not the only possibility (although it is the one we have examined
more closely). Another choice would be to include something like a
step function -- once the UV radiation is lower than a certain amount,
the rate of cooling increases greatly, but above the cutoff, it is
negligible. This idea of a critical density is essentially
the process advocated by Parravano and collaborators.  However, it can
be shown~\cite{car00} that there is little variation in the average
values of the components as the functional dependence of the cooling
is altered. In addition, the use of a step function in the cooling function can
lead to oscillatory behavior in the components that is
almost discontinuous and is therefore undesirable.

\subsection{Reaction-diffusion model}
\label{react-diff}

The system of equations $(\ref{eqns-one-zone})$ can be useful in 
understanding such things as the chemical evolution in a galaxy, but 
there already exists a substantial literature on one-zone models in 
galactic evolution. Our goal here is to work with a model, similar to 
those in chemical and biological non-equilibrium systems, where 
spatial and temporal patterns are generated. In these systems, there 
are different length scales: L$_{long}$ is the scale of the whole 
disk, the distance that UV radiation can travel; L$_{int}$ is the 
scale of distances between cloud complexes, and the distance core 
collapse is induced by supernovae; and L$_{short}$ is the scale of a 
single cloud, the distance new stars evaporate the GMC from which they 
condensed. Note that the reactions characterized by L$_{int}$ are 
catalytic, whereas those of L$_{long}$ and L$_{short}$ are inhibitive.

To incorporate the effects of these length scales and the 
inhomogeneities they produce, we introduce diffusion terms into the 
system of equations. Although diffusion is clearly appropriate for 
such things as the movement of material such as stars and clouds 
through the disk, it might seem more realistic to use a wave equation 
to represent the effects of shockwaves and radiation. However, as is 
well-known, there are wavelike solutions to reaction-diffusion 
equations, such as the prototype Fisher equation \cite{gri91}. 
Also, we can see from the success the cellular 
automata of Gerola, Seiden and Schulman \cite{sei-ger82} and the model of 
Elmegreen and Thomasson \cite{elm-tho92} that this is a valid 
approach.

Because the scale height of
the Milky Way is much smaller than that of the radius of the disk, we consider the model
on a two-dimensional annulus, leaving out the galactic bulge since it has little influence
on the star formation in the disk. Therefore the component functions 
we considered in Sec. \ref{one-zone} are now functions of the radius 
$\rho$ and the angle $\theta$, in addition to time $t$:
\[
c = c(\rho, \theta; t) \qquad g = g(\rho, \theta; t) \qquad s = s(\rho, 
\theta; t) \qquad r = r(\rho, \theta; t) \qquad h = h(\rho, \theta; t)
\]
Note that the radius and angle, along with the angular velocity $\omega(\rho)$ in the 
disk, will be the only Greek letters that are not parameters 
of the model. Being in a rotating system, with 
angular velocity $\omega(\rho)$, we must use the convective derivative, so 
that the evolution of the components is given by
\begin{mathletters}
    \label{eqns-react-diff}
\begin{eqnarray}
\frac{\partial c}{\partial t} + \omega(\rho) \frac{\partial c}{\partial \theta} &=& \frac{\alpha_1 g^2}{r} - \alpha_2 cr - \beta_1 ch - \gamma cs -
	\epsilon_{1} c^2\\
\frac{\partial g}{\partial t} + \omega(\rho) \frac{\partial g}{\partial \theta} &=& - \frac{\alpha_1 g^2}{r} + \alpha_2 cr + s + \gamma cs + \epsilon_{2} c^2
	+ \delta \\
\frac{\partial s}{\partial t} + \omega(\rho) \frac{\partial s}{\partial \theta} &=& \beta_{2} ch + \epsilon_{3} c^2 - s + D_s \nabla^2 s\\
\frac{\partial r}{\partial t} + \omega(\rho) \frac{\partial r}{\partial \theta} &=& \eta_1 s - \phi_{1} (c, g, r) + D_r \nabla^2 r\\
\frac{\partial h}{\partial t} + \omega(\rho) \frac{\partial h}{\partial \theta} &=& \eta_2 s - \phi_{2} (c, g, h) + D_s \nabla^2 h
\end{eqnarray}
\end{mathletters}
where the parameters have the same meaning as before, in the one-zone 
model.

\section{Numerical results and conclusions}

Now that we have our model, we want to know if there are any structures of
the right size that will arise. To do this, we consider a
linearization of the model, and see if there are any instabilities. 
We assume that the galaxy is two-dimensional to simplify the analysis, and
expand equations $(\ref{eqns-react-diff})$ to linear order around the steady state
(i.e. $c = {C_0} + C + \cdots$, and similarly for the other functions). 
We assume instabilities with a wave vector ${\bf k}$ growing with time scale
$\lambda$, so that, for example, $C = C_k e^{\lambda t} \cos ({\bf k \cdot
x})$. Note that we can find eigenfunctions of the differential operator made 
up of the Laplacian and the convective derivative 
\cite{car00,neu-fei88}. Then these equations give us a matrix equation of the
form $M^a_b v^b = \lambda v^a$, where $v^a$ is a column vector made
up of the component functions -- $v^a = (C,S,G,R,H)$. We can solve this
matrix equation for the eigenvalues $\lambda$, and find which modes of
instability are likely to grow exponentially as a function of the parameters.

To carry out the linearization analysis, we pick an arbitrary set of
parameters lying within the physical ranges, as described in Sec.
\ref{react-diff}. If we look at the maximum real eigenvalue (MRE), we
get a sense of which unstable modes will grow at the fastest rate.
In Figure 2, we see a graph of the MRE of the matrix $M^a_b$ as a function
of the logarithm of the wavenumber $k = |{\bf k}|$,
using the values
\[
\alpha_1 = \beta_1 = \eta_{1}/10 = 10 \eta_{2} = 1.0 \quad \alpha_2 = 
0.005 \quad \beta_2 = 0.9 \quad \gamma = 0.5
\]
\[
\delta = 0.002 \quad \epsilon_1 = 1.0 \quad 
\phi_1 = 0.6 \quad \phi_2 = 0.4 \quad
\epsilon_2 = 0.9 \quad \epsilon_3 = 0.04
\]
and the three diffusion constants
\[
D_h = 1 \qquad D_s = 10^{-3} \qquad D_r = 10^5
\]
Since we are using
$L_{\rm{int}} (\sim 10^2$ parsecs) as our length scale, then we note 
several things about the value of the MRE. It is always 
negative -- all modes are stable and will eventually decay to 
equilibrium, although some decay slower. These modes can 
possibly allow the formation of some spiral structure in the disk, 
even if it is not a permanent pattern. Note that the slowly decaying modes 
have pattern sizes of between $10^{-2} L_{\rm{int}}$ to $10^{2} L_{\rm{int}}$
(1 to $10^{4}$ parsecs). Unfortunately, this seems to be generic 
within the model, although there is the possibility that there 
exist parameter sets with positive MREs that we have not found.

Once we had a set of parameters that, at least for some time, would 
form structures of the right size, the equations $(\ref{eqns-react-diff})$ were 
numerically simulated using finite differencing, with the time evolution
given by operator splitting. Because the diffusion constant $D_r$ used in
the linearized analysis is so much larger than $D_h$ and $D_s$, it was
decided to use the mean field approximation for the UV radiation: the
radiation was spread instantaneously across the galactic disk between time
steps, instead of diffusing. Because the time increment used in the
simulation was on the order of the light crossing time of the disk (about
$10^4$ years), this is not too unphysical a proposition. To take into
account the effects of gravity, a constant linear velocity $v$ was given to
all the material, approximating the situation in the Milky Way. This was
done by using $\omega = v / r$ in the convective derivative.

We present a picture of our results in Figure 3.
The initial data is given on an annulus with $r_{\rm{min}}
= 50 L_{\rm{int}}\ (5 \times 10^3$ parsecs) and $r_{\rm{max}} = 150L_{\rm{int}}\
(1.5 \times 10^4 $ parsecs), and is just a gaussian ``blip'' for one 
component ($w$), at an arbitrarily choosen location. 
The disk is rotating with a constant linear speed $v = 
30 L_{\rm{int}} / \tau$, which is the velocity of the Sun in the Milky Way. 
The boundary conditions on the inner and outer edge of the annulus is such that
the radial derivatives of all the components are zero, i.e. there is no flux through
the edges. This comes from the rationale that there are no clouds or stars
infalling from the outside, and the central bulge does not 
interact much with the annulus. The run covered a time of about $ 200 \tau$, a timespan of two billion 
years or approximately eight revolutions of the galaxy. As one can see from the 
figures, spiral structure indeed is formed. The spirals develop and grow 
in length as they rotate with the galactic disk. One could imagine with several such 
initial gaussian ``blips'', several spiral arms would form, thus resembling a real,
flocculant, spiral galaxy. 

However, after about $100 \tau$ or one billion years, the ends of a spiral arm, begin to
interact (diffuse) with each other and eventually (at about $1.5$ billion years) merge 
forming a ring like structure. 
This appears to be a generic limitation of the model. Several initial data 
and parameter sets were
tried with the same results. It seems that only a modification of the model will be
able to save the spirals from this eventual, unfortunate fate! Work on this issue is 
currently underway and shall be presented elsewhere. Even with this limitation, this
model competes well with other current models. 

Beyond this issue, there are a number of future directions that this work can be taken. 
This run only used one specific set of initial data; one can hope that spiral formation
does not depend strongly on initial conditions. We have made several runs
with differing initial data, but we have not further explored this area.
We are also lacking a detailed study of the effects of changes in the values of the 
parameters; because of the large number of parameters and components, 
however, it is difficult to see how an exhaustive search for a more 
fruitful choice could be made. Modifying the boundary conditions is another avenue to consider -- for
example, allowing radiation to flow in from outside the galaxy, or to have
interaction between the disk and the central bulge. Also, we might alter the
role of the UV radiation in the model. One consideration is that we have not
given the radiation any type of radial fall-off as it leaves the source, but
instead used a mean field approximation. This, unfortunately, heats the
entire disk evenly. If, instead, there is a great deal of radiation in the
space near the source, and little far away, this might give a more realistic
feedback mechanism to inhibit star formation near the source but allow it
further away. There are, of course, other ways that the model can be changed
to make it more realistic, such as adding in more reaction mechanisms. 

\section{Acknowledgements}

The authors would like to thank Lee Smolin for introducing us to this subject,
as well as many helpful discussions. Also, we appreciate the help of Christopher
Beetle with numerical methods, Olaf Dreyer with Mathematica and Sameer Gupta
with \LaTeX. Most of this work was done when the authors were at the Center for Gravitational Physics and 
Geometry at the Pennsylvania State University. 

One of us (GK) wishes to thank Southampton College of Long Island University for 
research support and computational facilities. 

\begin{figure}                                                           
\label{flow}                                                     
\centerline{\psfig{figure=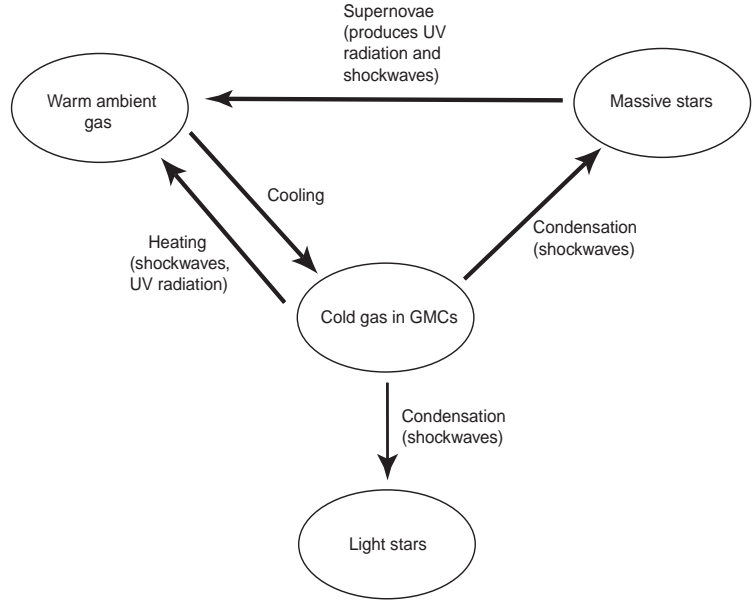,width=100mm,height=80mm}}                                  
\caption{Diagram indicating matter flows in galactic disks, along with   
their catalysts.}                                                        
\end{figure}

\narrowtext

\begin{table}
    \caption{Description of the various components of the model; in 
    the numerical simulation, these are all functions of radius, 
    angle and time.}
    \label{table1}
\begin{tabular}{cl}
{\bf Variable} & {\bf Description} \\
\tableline
$c$ & Cold gas in GMCs \\
$g$ & Warm, ambient gas \\
$s$ & Massive stars \\
$d$ & Light stars \\
$r$ & Density of UV radiation \\
$h$ & Density of shockwaves from supernovae \\
\end{tabular}
\end{table}

\mediumtext

\begin{table}
\caption{Parameters of the model; the rationale behind these choices 
           is given in [7]. }
\label{table2}
\begin{tabular}{ccl}
\hline
		&		& \\
{\bf Parameter} & {\bf Value} 	& {\bf Description} \\
\hline
		&		& \\
$\alpha_1$	& 0.1 - 1	& Rate of GMC increase via cooling\\
$\alpha_2$	& $5 \times 10^{-2} - 10^{-4}$	& Rate of GMC decrease via heating\\
$\beta_1$	& 0.1 - 10	& GMC destruction rate by shockwaves\\
$\beta_2$	& 0.02 - 2	& Massive star production rate by shockwaves\\
$\beta_3$	& 0.08 - 8	& Light star production rate by shockwaves\\
$\gamma$	& $3 \times 10^{-2} - 5$ & GMC destruction rate by massive star heating\\
$\epsilon_1$	& 0.2 - 2	& Destruction rate of clouds via collisions \\
$\epsilon_2$ 	& 0.18 - 1.8	& Formation rate of warm gas from cloud-cloud collisions \\
$\epsilon_3$	& 0.004 - 0.04	& Formation rate of massive stars from cloud-cloud collisions\\
$\epsilon_4$	& 0.02 - 0.2	& Formation rate of massive stars from cloud-cloud collisions\\
$\eta_1$	& 10		& Production rate of UV radiation by SNe \\
$\eta_2$	& 0.02 - 1	& Production rate of shockwaves of SNe \\
$\phi_1$	& 0.6		& Average optical depth of UV radiation \\
$\phi_2$	& 0.4		& Average ``shock depth'' of SNe shockwaves \\
$\delta$	& 0.001 - 0.003	& Rate of warm gas accretion onto the galactic disk \\
\hline
\end{tabular}
\end{table}

\begin{figure}                                                           
\label{linear_graph}                                                     
\centerline{\psfig{figure=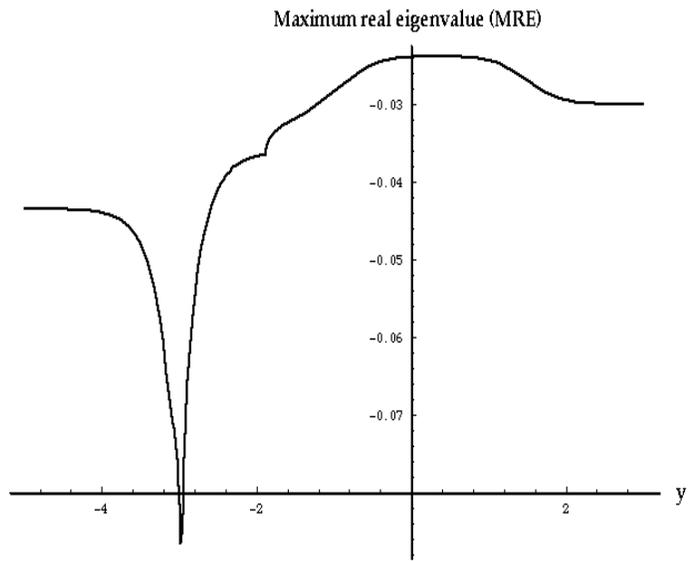,width=100mm,height=75mm}}            
\caption{A graph of the maximum real eigenvalue (MRE) versus             
$y=\log({\bf k})$.}                                                      
\end{figure}                                                             

\begin{figure}
\centerline{\psfig{figure=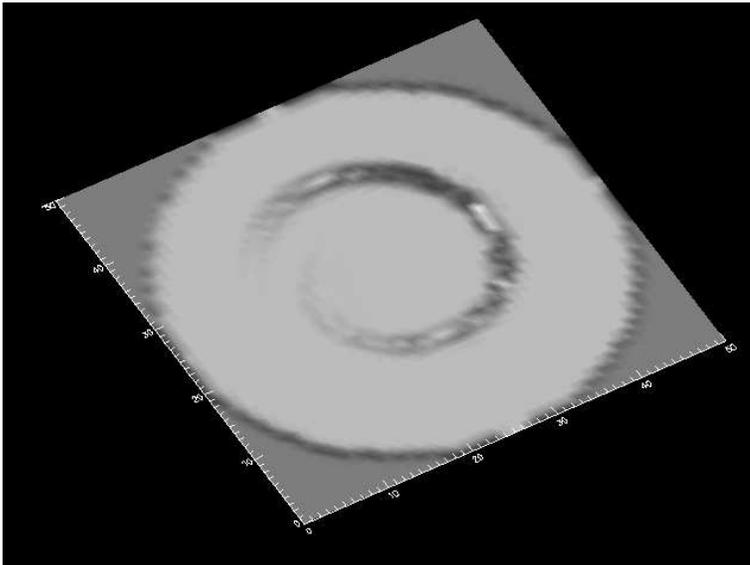,width=100mm,height=75mm}}
\caption{A surface plot of the (light) stars component at $250$ million years, about one revolution of the galactic disk. 
Note a spiral arm forming and developing.}
\end{figure}

\end{document}